\documentclass[conference]{IEEEtran}
\IEEEoverridecommandlockouts
\usepackage{cite}
\usepackage{amsmath,amssymb,amsfonts}
\usepackage{algorithmic}
\usepackage{graphicx}
\usepackage{textcomp}
\usepackage{xcolor}
\usepackage{multirow}
\usepackage[export]{adjustbox}[2011/08/13]\usepackage[export]{adjustbox}[2011/08/13]

\def\BibTeX{{\rm B\kern-.05em{\sc i\kern-.025em b}\kern-.08em
    T\kern-.1667em\lower.7ex\hbox{E}\kern-.125emX}}
\begin{document}

\title{Hybrid Swin Deformable Attention U-Net for Medical Image Segmentation\\}

\author{\IEEEauthorblockN{1\textsuperscript{st} Lichao Wang$^*$}
\IEEEauthorblockA{\textit{Department of Computing} \\
\textit{Imperial College London}\\
London, UK\\
l.wang22@imperial.ac.uk}
\and
\IEEEauthorblockN{2\textsuperscript{nd} Jiahao Huang$^*$}
\IEEEauthorblockA{\textit{National Heart and Lung Institute} \\
\textit{Imperial College London}\\
London, UK \\
 j.huang21@imperial.ac.uk}
\and
\IEEEauthorblockN{3\textsuperscript{rd} Xiaodan Xing}
\IEEEauthorblockA{\textit{National Heart and Lung Institute} \\
\textit{Imperial College London}\\
London, UK \\
x.xing@imperial.ac.uk}
\and
\IEEEauthorblockN{4\textsuperscript{th} Guang Yang}
\IEEEauthorblockA{\textit{National Heart and Lung Institute, Imperial-X} \\
\textit{Imperial College London}\\
London, UK \\
g.yang@imperial.ac.uk}
}

\maketitle
\renewcommand{\thefootnote}{\fnsymbol{footnote}} 
\footnotetext[1]{These authors contributed equally to this work.} 

\begin{abstract}
Medical image segmentation is a crucial task in the field of medical image analysis. Harmonizing the convolution and multi-head self-attention mechanism is a recent research focus in this field, with various combination methods proposed. However, the lack of interpretability of these hybrid models remains a common pitfall, limiting their practical application in clinical scenarios. To address this issue, we propose to incorporate the Shifted Window (Swin) Deformable Attention into a hybrid architecture to improve segmentation performance while ensuring explainability. Our proposed Swin Deformable Attention Hybrid UNet (SDAH-UNet) demonstrates state-of-the-art performance on both anatomical and lesion segmentation tasks. Moreover, we provide a direct and visual explanation of the model focalization and how the model forms it, enabling clinicians to better understand and trust the decision of the model. Our approach could be a promising solution to the challenge of developing accurate and interpretable medical image segmentation models. (Our code can be made open source, once accepted.)
\end{abstract}


\section{Introduction}
Accurate medical image segmentation is crucial in many clinical scenarios, and interpretability is essential for segmentation models as it enables clinicians to understand how the model arrives at its decision and provides insight into its working mechanism. However, the feature maps embedded in segmentation networks are often difficult to interpret due to their indirect correlation with the network optimization procedure. Therefore, segmentation methods that can simultaneously provide interpretation masks are required.

Gradient-assisted methods such as Gradient-weighted Class Activation Mapping (Grad-CAM) \cite{Grad-CAM} and Layer-Wise Relevance Propagation \cite{Bach2015OnPE} have been proposed as major solutions for interpretation visualization. 
However, these methods do not adequately capture fine-grained information regarding object boundaries, and as external extensions, they do not offer an intrinsic understanding of the decision-making process. The introduction of the multi-head self-attention (MSA) mechanism has partially mitigated this limitation \cite{TransUNet,LeViT-UNet,Hatamizadeh2021UNETRTF,SwinE-Net,HybridCTrm,UNet2022,convswinUNet}. These methods not only provide attention score heatmaps that enable internal interpretability of the model but also improve segmentation accuracy through the data-specific nature of the MSA \cite{How-vit-works}.

\begin{figure}[hbt]
    \vspace{-2ex}
    \centering
    \includegraphics[width=0.46\textwidth]{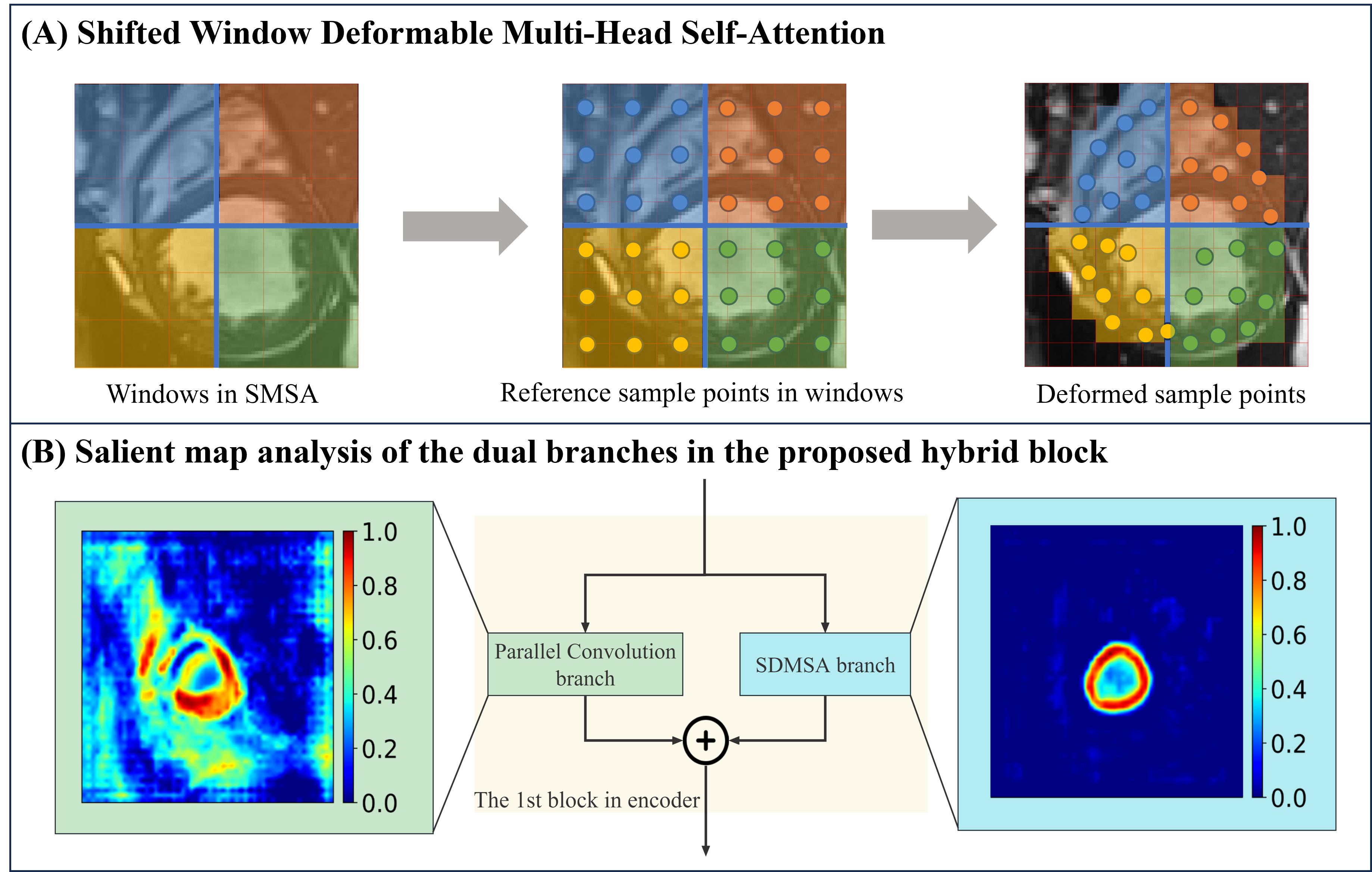}
    \vspace{-1ex}
    \caption{Description and analysis of the proposed hybrid block with the Swin Deformable Multi-Head Self-Attention (SDMSA) module and Parallel Convolution. 
    (A) the deformation of the sample points in the SDMSA module;
    (B) the salient maps of the Parallel Convolution branch and the SDMSA branch, captured by the SEG-GRAD-CAM \cite{Vinogradova2020TowardsIS}, with myocardium as the segmentation target.
    }
    \label{fig:intro}

    \vspace{-1ex}
\end{figure}

Despite the advantages of the attention mechanism, redundant attention sometimes occurs due to the lack of deformability of conventional attention modules \cite{xia2022vision}. For example, in a left ventricle (LV) walls segmentation task (Fig. \ref{fig:intro} (A)), the Shifted Window (Swin) multi-head self-attention (SMSA) branch redundantly computes the MSA on all patches in each window due to the lack of ability to capture the complicated, irregularly distributed, and deformed LV wall area. This redundant attention raises questions regarding the interpretation and increases the computational cost.

To address this limitation, we develop a novel hybrid block containing a Swin Deformable MSA (SDMSA) module (Fig. \ref{fig:intro} (A)) which enables the MSA mechanism to precisely focus on the segmentation target and remove redundancy. 
However, models with only the SDMSA in its basic block can result in the loss of detailed information, due to the patch-based mechanism and the lack of the local prior. Therefore, a computation-efficient convolution branch in parallel is designed to acquire detailed texture features \cite{How-vit-works}. As shown in Fig. \ref{fig:intro} (B), the Parallel Convolution branch focuses more on the detailed features, while the SDMSA branch concentrates on the holistic shape structure.

We hypothesize that 
(1) the SDMSA module enables the model to focus more precisely on the segmentation targets and provides explainability; 
(2) our proposed model can achieve state-of-the-art performance on multiple medical image segmentation datasets, incorporating different clinical questions. 
The main contributions are listed as follows: 
(1) we propose a Swin Deformable Attention Hybrid UNet (SDAH-UNet) model for medical image segmentation; 
(2) our SDMSA with Parallel Convolution (SDAPC) block improves the segmentation performance based on the SDMSA module, which strengthens the focus of the model on the segmentation targets and can provide explanations; 
(3) in the experiment section, SDAH-UNet significantly outperforms state-of-the-art models, on automatic cardiac diagnosis challenge (ACDC) \cite{Bernard2018DeepLT}, and brain tumor segmentation 2020 (BraTS2020) \cite{Menze2015TheMB}.

\section{Methodology}

\begin{figure}[h]
    \vspace{-2ex}
    \centering
    \includegraphics[width=0.48\textwidth]{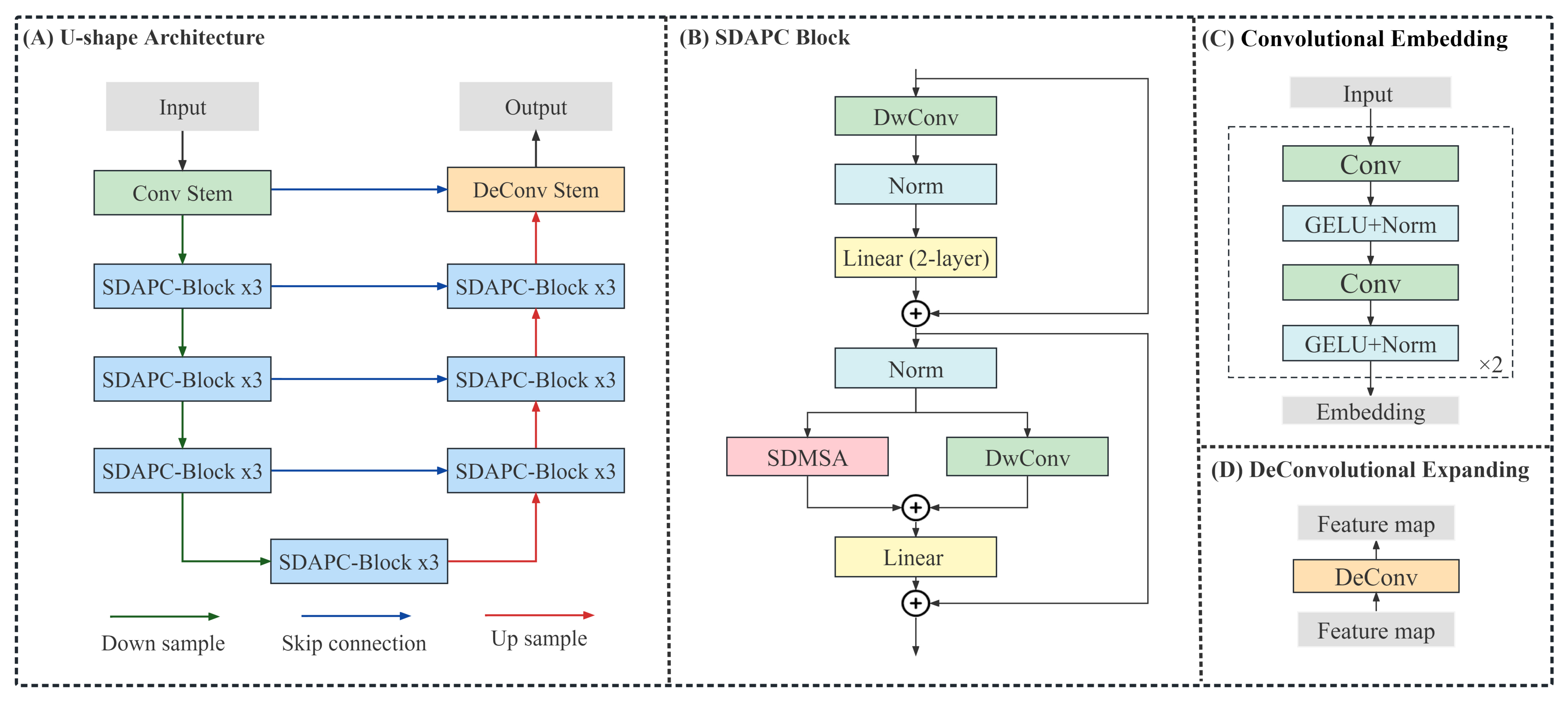}
    \vspace{-1ex}
    \caption{The illustration of Swin Deformable Attention Hybrid UNet (SDAH-UNet). 
    (A) the overall architecture of SDAH-UNet; 
    (B) the proposed SDMSA with Parallel Convolution (SDAPC) block;
    (C) the structure of the Convolutional Embedding module;
    (D) the structure of the DeConvlutional Expanding module.}
    \label{fig:model}
    \vspace{-2ex}
\end{figure}

\subsection{U-shape Architecture}

Our proposed model is an end-to-end model that takes an image slice as input and produces a corresponding segmentation mask as output. The proposed model follows a symmetrical design paradigm, with both the encoder and decoder consisting of three SDAPC blocks. A convolutional stem is added at the beginning of the encoder, and a deconvolutional stem is added at the end of the decoder. An additional SDAPC block is inserted in the bottleneck (Fig. \ref{fig:model} (A)).

\subsection{Convolutional Embedding Module and DeConvolutional Expanding Module}

Drawing inspiration from recent studies \cite{nnFormer,UNet2022}, we directly adopt successive convolutional layers with small kernels as the embedding module. This design aims to precisely encode pixel-level spatial information while minimizing computational complexity and maintaining a large receptive field. The embedding module comprises four convolutional layers, each followed by a Gaussian Error Linear Unit (GELU) \cite{hendrycks2016gelu} activation function and a layer normalization layer. The expanding module is comprised of a deconvolutional layer (Fig. \ref{fig:model} (C) and (D)). This simple yet effective architecture can be readily extended to support higher-resolution inputs.

\subsection{Swin Deformable Multi-Head Self-Attention with Parallel Convolution Block}

Convolution and MSA have been shown to have distinct weighting mechanisms \cite{UNet2022}, filtering effects \cite{How-vit-works}, and levels of focus on feature map details, which can be complementary to each other. To leverage the benefits of both mechanisms, our SDAPC block is designed in a parallel paradigm (Fig. \ref{fig:model} (B)). Partitioned by the two residual connections, the SDAPC block comprises two divisions. In the first division, the input $X$ first undergoes a Depthwise Convolution (DwConv) \cite{chollet2017xception} layer, ${\rm DwConv}(\cdot)$, followed by a standard 2-layer multi-layer perceptron (MLP), which can be described as
\begin{equation}
    \bar{X} = {\rm FC_2}({\rm GELU}({\rm FC_1}({\rm LN}({\rm DwConv}(X))))) + X,
    \label{eq:division1}
\end{equation}
in which ${\rm FC_1}(\cdot)$ and ${\rm FC_2}(\cdot)$ are full connection layers. ${\rm GELU}(\cdot)$ and ${\rm LN}(\cdot)$ denote the GELU and Layer Normalization layers.

In the second division, $\bar{X}$ is passed to a paralleled DwConv layer, ${\rm DwConv}(\cdot)$, and a SDMSA layer, after which their outputs are concatenated and passed through the final ${\rm FC}$ layer. Inspired by the design of ConvNeXt \cite{Liu2022ConvNeXt}, the kernel size of DwConv is set as 7. The process can be described as
\begin{equation}
    \hat{X} ={\rm FC}({\rm SDMSA}({\rm LN}(\bar{X})) + {\rm DwConv}({\rm LN}(\bar{X}))) + \bar{X}.
    \label{eq:division2}
\end{equation}

\subsection{Shifted Window Deformable Multi-Head Self-Attention}

Followed by the Deformation Attention Transformer \cite{xia2022vision} and the Swin Deformable Attention U-Net Transformer \cite{Huang2022SwinDA}, we introduce the deformation operation into the SMSA, resulting in the SDMSA. The input feature map $F\in \mathbb{R}^{N \times W \times H}$ is first split into $N_w$ non-overlapping windows and then divided into $N_h$ heads along the channels. After the partition, $N_w \times N_h$ sub-feature maps are generated, where $N_w = WH/W_s^2$, $W_s$ stands for window size. For a sub-feature map $F_{i}^{j}$, where $i$ and $j$ represent the $i^{\rm th}$ window and the $j^{\rm th}$ head respectively, we calculate the query $q_i^j$, and set up a group of uniformly distributed reference sample points $p_i^j$. The reference sample points are added with the offsets $\Delta p_i^j$ obtained from a two-layer convolution neural network ${\rm CNN_{offest}(\cdot)}$, to generate the deformed sample points $p_i^j + \Delta p_i^j$. A bilinear interpolation function $\phi(\cdot)$ is used to generate the sampled features. Then keys and values are calculated based on the sampled features. The SDMSA can be presented as follows

\begin{equation}
    q_i^j = F_i^j {W_{Q}}^j_i,\quad \Delta p_i^j = {\rm CNN_{offest}}(q_i^j),
    \label{eq:kqv}
\end{equation}
\vspace{-4ex}


\begin{equation}
    \hat{F_i^j} = {\rm \phi} (F_i^j; p_i^j+\Delta p_i^j)
    \label{eq:offset}
\end{equation}
\vspace{-4ex}

\begin{equation}
     k_i^j = \hat{F_i^j}{W_{K}}^j_i, \quad v_i^j = \hat{F_i^j}{W_{V}}^j_i,
    \label{eq:offset_2}
\end{equation}
\vspace{-4ex}

\begin{equation}
    Z_i^j = {\rm SoftMax}(q_i^jk_i^j/\sqrt{d}+b_i^j)v_i^j,
    \label{eq:Attention}
\end{equation}

\noindent where $Wq_i^j$, $Wk_i^j$, and $Wv_i^j$ are the projection matrices for queries, keys, and values, respectively. $b_i^j$ represents the fixed positional bias \cite{swin-Transformer}. The dimension of each head is represented as $d = C/N_h$, where $C$ is the number of output channels and $N_h$ is the number of heads.

\section{Experiments}

Experiments were conducted on two medical image segmentation tasks, i.e., ACDC \cite{Bernard2018DeepLT} and BraTS2020 \cite{Menze2015TheMB}, which represent different clinical questions and MRI sequences to validate the efficacy of our proposed methods. The ACDC dataset is an anatomical segmentation task, with single MRI modality serving as the input. BraTS2020 is a lesion segmentation task, with multi-modal MRI data as the input. MRI volumetric data was split into slices for the model training.

\subsection{Implementation Details and Evaluation Methods}
We conducted our experiments on an NVIDIA RTX3090 GPU with 24GB GPU RAM. Dice loss $\mathcal{L}_{\rm dice}$ and cross-entropy loss $\mathcal{L}_{\rm CE}$ were combined as the loss function. 

During the training process, the batch size and input resolution were set to 24 and $224\times224$ for both datasets. The initial learning rate was set to $2\times10^{-4}$ and decayed every 10,000 steps by 0.5 from the $50,000^{\rm th}$ step. During the inference stage, our model makes predictions using a sliding window approach on both datasets\cite{Isensee2018}. We set 112, 0.5 times of the crop size (224), as the step size. For patch aggregation, the Gaussian importance weighting strategy was utilized, giving more weight to pixels in the center area.

The evaluation of performance in this study incorporates the utilization of two metrics: the Dice similarity coefficient (DSC) \cite{dice1945measures} and the Hausdorff95 metric (HD95) \cite{birsan2006one}. The statistical analysis conducted to ascertain the significance of the results involves the application of a paired sample t-test. For the purpose of evaluating the model size and time complexity, two quantitative measures were employed: the count of parameters (Params) and the rate of floating-point operations per second (FLOPs).

\subsection{Datasets}

The ACDC data \cite{Bernard2018DeepLT} is a collection of 100 cine MRI scans (1,902 slices). The data was randomly split into training, validation, and test sets. Within the ACDC dataset, short axis cine MRI scans are available for each patient, consisting of 12-35 frames, along with end-diastole (ED) and end-systole (ES) frames. The dataset provides segmentation targets for three foreground classes: right ventricle (RV), myocardium (MYO), and left ventricle (LV).

The BraTS2020 data \cite{Menze2015TheMB} consists of 369 cases (57,195 slices) of brain MRI scans, each with four modalities: native T1-weighted, post-contrast T1-weighted, T2-weighted, and Fluid Attenuated Inversion Recovery scans. The data was randomly split into training, validation, and test sets. The segmentation accuracy was evaluated for the enhancing tumor region (ET, label 1), regions of the tumor core (TC, labels 1 and 4), and the whole tumor region (WT, labels 1, 2, and 4).

\subsection{The State-Of-The-Art Segmentation Performance}
The quantitative results of the comparison on two general medical image segmentation datasets are reported in Table \ref{tab:ACDC} and \ref{tab:brats}. Results demonstrate that the proposed SDAH-UNet, with strong generalization ability, can be applied to multiple datasets including anatomical and lesion segmentation datasets. Table \ref{tab:flops} demonstrates that SDAH-UNet exhibits a marginal increase in parameter count (attributed to CNN$_{offset}$ and Parallel Convolution) compared to other SMSA-based models (SwinUNet \cite{Cao2021}, UNet-2022), while maintaining relatively lower FLOPs. This suggests that SDAH-UNet achieves state-of-the-art performance with a minimal increment of computational overhead.

\begin{table}[htb]\label{tab:segmentation}
\centering
\scalebox{0.8}{
\begin{tabular}{l|cc|clclcl}
\hline
\multirow{2}{*}{Methods} & \multicolumn{1}{c|}{DSC} & \multicolumn{1}{c|}{HD95} & \multicolumn{6}{c}{DSC}                                                     \\ \cline{2-9} 
                         & \multicolumn{2}{c|}{AVG}                             & \multicolumn{2}{c|}{RV} & \multicolumn{2}{c|}{MYO} & \multicolumn{2}{c}{LV} \\ \hline
UNet   & 87.68 & 1.55 & \multicolumn{2}{c}{83.82 $\pm$ 11.04$^\dagger$} & \multicolumn{2}{c}{84.21 $\pm$ 9.97$^\dagger$} & \multicolumn{2}{c}{95.04 $\pm$ 5.09$^\dagger$} \\
nnUNet    & \underline{91.86} & \underline{1.22} & \multicolumn{2}{c}{\underline{88.86 $\pm$ 7.43}} & \multicolumn{2}{c}{\underline{89.77 $\pm$ 2.29}} & \multicolumn{2}{c}{\textbf{96.97 $\pm$ 0.75}} \\
SwinUNet   & 88.08 & 1.61 & \multicolumn{2}{c}{85.24 $\pm$ 8.48$^\dagger$} & \multicolumn{2}{c}{84.45 $\pm$ 7.76$^\dagger$} & \multicolumn{2}{c}{94.56 $\pm$ 4.64$^\dagger$} \\
TransUNet & 89.48 & 1.41 & \multicolumn{2}{c}{87.21 $\pm$ 9.23$^\dagger$} & \multicolumn{2}{c}{85.73 $\pm$ 6.01$^\dagger$} & \multicolumn{2}{c}{95.52 $\pm$ 3.39$^\dagger$} \\ 
UNet-2022 & 90.42 & 1.41 & \multicolumn{2}{c}{87.36 $\pm$ 9.09$^\dagger$} & \multicolumn{2}{c}{87.78 $\pm$ 5.09$^\dagger$} & \multicolumn{2}{c}{96.13 $\pm$ 3.28$^\dagger$} \\ \hline
SDAH-UNet & \textbf{92.23} & \textbf{1.22} & \multicolumn{2}{c}{\textbf{89.78 $\pm$ 9.97}} & \multicolumn{2}{c}{\textbf{90.02 $\pm$ 4.23}} & \multicolumn{2}{c}{\underline{96.91 $\pm$ 1.91}} \\ \hline
\end{tabular}
}
\vspace{1.2ex}
\caption{Comparison with state-of-the-art models on the ACDC dataset. $^\dagger$: p $<$ 0.05 using paired sample t-test compared with our proposed SDAH-UNet. The bolded entries in the table represent the best results, while the underlined entries indicate the second-best results. RV: Right Ventricle, MYO: Myocardium, LV: Left Ventricle. }
\label{tab:ACDC}
\end{table}

\begin{table}[htb]

\centering
\scalebox{0.8}{
\begin{tabular}{l|cc|clclcl}
\hline
\multirow{2}{*}{Methods} & \multicolumn{1}{c|}{DSC} & \multicolumn{1}{c|}{HD95} & \multicolumn{6}{c}{DSC}                                                     \\ \cline{2-9} 
                         & \multicolumn{2}{c|}{AVG}                             & \multicolumn{2}{c|}{WT} & \multicolumn{2}{c|}{TC} & \multicolumn{2}{c}{ET} \\ \hline
UNet   & 82.27 & 9.04 & \multicolumn{2}{c}{85.68 $\pm$ 8.82$^\dagger$} & \multicolumn{2}{c}{85.46 $\pm$ 8.67$^\dagger$} & \multicolumn{2}{c}{75.69 $\pm$ 8.92$^\dagger$} \\
nnUNet   & 83.64 & \underline{4.30} & \multicolumn{2}{c}{86.45 $\pm$ 10.89$^\dagger$} & \multicolumn{2}{c}{82.35 $\pm$ 7.28$^\dagger$} & \multicolumn{2}{c}{\underline{82.13 $\pm$ 11.37}$^\dagger$} \\
SwinUNet   & 84.56 & 9.25 & \multicolumn{2}{c}{87.47 $\pm$ 7.10$^\dagger$} & \multicolumn{2}{c}{85.50 $\pm$ 9.08$^\dagger$} & \multicolumn{2}{c}{80.71 $\pm$ 8.21$^\dagger$} \\
TransUNet & \underline{84.64} & 6.02 & \multicolumn{2}{c}{\underline{87.87 $\pm$ 9.03}$^\dagger$} & \multicolumn{2}{c}{\underline{85.88 $\pm$ 
9.73}$^\dagger$} & \multicolumn{2}{c}{80.19 $\pm$ 8.88$^\dagger$} \\ 
UNet-2022 & \underline{84.76} & 5.52 & \multicolumn{2}{c}{\underline{87.95 $\pm$ 9.03}$^\dagger$} & \multicolumn{2}{c}{\underline{85.97 $\pm$ 
9.61}$^\dagger$} & \multicolumn{2}{c}{80.35 $\pm$ 8.68$^\dagger$} \\ \hline
SDAH-UNet & \textbf{86.90} & \textbf{3.65} & \multicolumn{2}{c}{\textbf{90.21 $\pm$ 3.55}} & \multicolumn{2}{c}{\textbf{88.13 $\pm$ 5.23}} & \multicolumn{2}{c}{\textbf{82.37 $\pm$ 5.45}} \\ \hline
\end{tabular}
} 
\vspace{1.2ex}
\caption{Comparison with state-of-the-art models on the BraTS2020 dataset. $^\dagger$: p $<$ 0.05 using paired sample t-test compared with our proposed SDAH-UNet. The bolded entries in the table represent the best results, while the underlined entries indicate the second-best results. ET: Enhanced Tumor, TC: Tumor Core, WT: Whole Tumor.}
\label{tab:brats}
\vspace{-1ex}
\end{table}

\begin{table}[!h]
\scalebox{0.8}{
\begin{tabular}{c|ccclll}
\hline
Model                      & UNet                        & nnUNet                      & SwinUNet                    & TransUNet & UNet-2022 & SDAH-UNet \\ \hline
Params                     & 7.852M                      & 10.468M                     & 155.384M                    & 312.809M  & 162.894M  & 162.942M\\ \hline
\multicolumn{1}{l|}{FLOPs} & \multicolumn{1}{l}{10.772G} & \multicolumn{1}{l}{31.340G} & \multicolumn{1}{l}{33.201G} & 53.410G   & 37.973G   & 37.950G   \\ \hline
\end{tabular}
}
\label{tab:flops}
\vspace{1ex}
\caption{Number of parameters (Params) and time complexity (FLOPs) comparison}
\vspace{-1ex}
\end{table}

\subsection{The Accurate Interpretation}
\begin{figure}[htb]
    \vspace{-3ex}
    \includegraphics[width=0.5\textwidth]{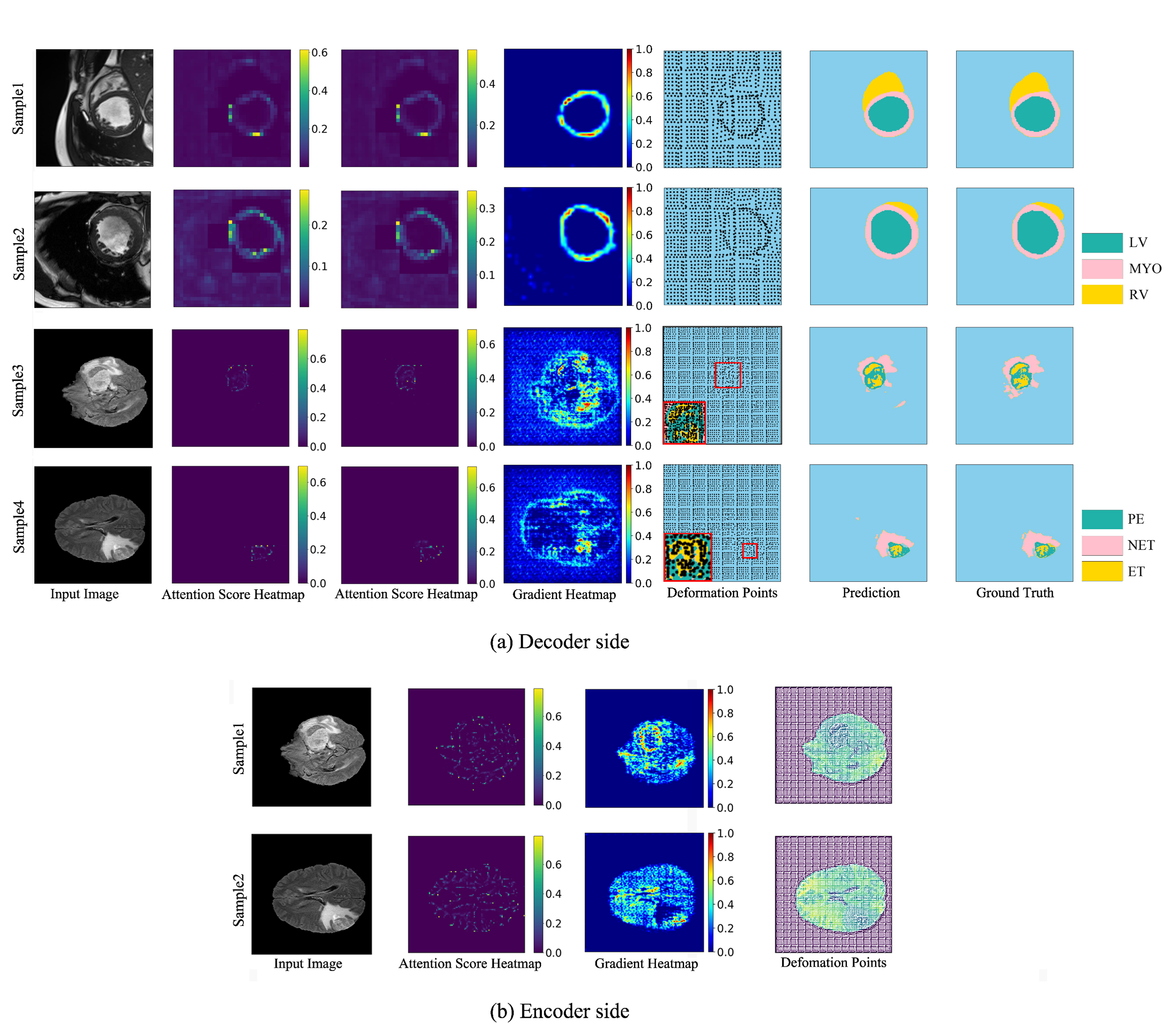}
    \vspace{-4ex}
    \caption{Visual explanation for SDAH-UNet. For the decoder, the Attention Score Heatmap, Gradient Heatmap, Deformation Points, and Deformation Field are captured from the last block, with down-sampling process for the deformation points. For the encoder, all figures are captured from the first block, without down-sampling process. LV, MYO, and RV stand for left ventricle, myocardium, and right ventricle; PE, NET, and ET stand for peritumoral edema, necrotic and non-enhancing tumor, and enhancing tumor.}
    \label{fig:ALL_samples_dec}
    \vspace{-2.5ex}
\end{figure}

The visual explanations for decoder and encoder are shown in Fig.~\ref{fig:ALL_samples_dec}. For the ACDC dataset, all visual explanation methods were observed to effectively illustrate the focalization of the decoder. Although the deformation of sample points impaired the visualization of the feature map, the outline of the myocardium still remained discernible. In the case of the more challenging BraTS2020 dataset, the attention score heatmap only provided a general indication of the tumor location, and the gradient heatmap had excessive highlight areas. 
In contrast, the utilization of deformation points presented a clearer representation of the model focalization. Global area analysis supported this finding by revealing that only sample points sufficiently close to the tumor exhibited deformation and were clustered in the tumor area. Moreover, when examining the results in a zoomed-out window, it could be observed that sample points were concentrated in the enhanced tumor region, within the whole tumor area. In the encoder side, the attention score heatmap was found to only depict the approximate location of the tumor. The gradient heatmap has shown that, in sample 1, the model paid more attention to the highlighted area; however, for sample 2 the gradient heatmap has shown that the model ignores the highlighted tumor area. On the other hand, the deformation points illustrated that the model focused on both the highlighted areas and the boundaries.

This outcome supports our hypothesis that compared to other visual explanation methods, the deformation points can facilitate a better understanding of the focalization by offering a clearer and more robust indication of the model focalization. Moreover, it also indicates that, in the windows nearing the segmentation target, the MSA computation focuses more precisely on the information-dense area, which reduces the computation redundancy.  

\subsection{Ablation Studies}

We set two groups of ablation studies to investigate the influence of the number of the SDAPC blocks and the two branches of the SDAPC block. 

Table.~\ref{tab:ablation blocks} shows that the performance of the model improved when the N-block (block with SMSA module and Parallel Convolution) was gradually replaced by the SDAPC block. replacing the first N-block with the SDAPC block lead to nearly 1.2\% performance enhancement for DSC. However, continuing to replace the N-blocks could not have a significant influence. Table.~\ref{tab:ablation two branch} illustrates that removing each branch from the SDAPC block leads to a decline in performance, nearly 2\% DSC. 

\begin{table}[h]
\centering
\scalebox{0.9}{
\begin{tabular}{l|cc|clclcl}
\hline
\multirow{2}{*}{Methods} & \multicolumn{1}{c|}{DSC} & \multicolumn{1}{c|}{HD95} & \multicolumn{6}{c}{DSC}                                                     \\ \cline{2-9} 
                         & \multicolumn{2}{c|}{AVG}                            & \multicolumn{2}{c|}{RV} & \multicolumn{2}{c|}{MYO} & \multicolumn{2}{c}{LV} \\ \hline
NNNN & 90.42 & 1.41 & \multicolumn{2}{c}{87.36 $\pm$ 9.09} & \multicolumn{2}{c}{87.78 $\pm$ 5.09} & \multicolumn{2}{c}{96.13 $\pm$ 3.28} \\
DNNN & 91.61 & 1.22 & \multicolumn{2}{c}{89.30 $\pm$ 5.80} & \multicolumn{2}{c}{88.79 $\pm$ 2.82} & \multicolumn{2}{c}{96.74 $\pm$ 0.88} \\
DDNN & 91.86 & 1.23 & \multicolumn{2}{c}{89.87 $\pm$ 5.34} & \multicolumn{2}{c}{88.88 $\pm$ 3.10} & \multicolumn{2}{c}{96.83 $\pm$ 0.87} \\
DDDN & 91.99 & 1.22 & \multicolumn{2}{c}{\textbf{90.02 $\pm$ 6.21}} & \multicolumn{2}{c}{89.35 $\pm$ 3.01} & \multicolumn{2}{c}{96.61 $\pm$ 1.83} \\
DDDD & \textbf{92.23} & \textbf{1.22} & \multicolumn{2}{c}{89.78 $\pm$ 9.97} & \multicolumn{2}{c}{\textbf{90.02 $\pm$ 4.23}} & \multicolumn{2}{c}{\textbf{96.91 $\pm$ 1.91}} \\ \hline
\end{tabular}
}
\vspace{1ex}
\caption{Comparison of the different numbers of SDAPC blocks adopted in the model on the ACDC dataset. N stands for the the block with SMSA and Parallel Convolution (without deformation operation), and D stands for the SDAPC block (with deformation operation)}
\label{tab:ablation blocks}
\end{table}

\begin{table}[h]
\centering
\scalebox{0.9}{
\begin{tabular}{l|cc|clclcl}
\hline
\multirow{2}{*}{Branch Type} & \multicolumn{1}{c|}{DSC} & \multicolumn{1}{c|}{HD95} & \multicolumn{6}{c}{DSC}                                                     \\ \cline{2-9} 
                             & \multicolumn{2}{c|}{AVG}                            & \multicolumn{2}{c|}{RV} & \multicolumn{2}{c|}{MYO} & \multicolumn{2}{c}{LV} \\ \hline
SDMSA-branch & 90.48 & 1.45 & \multicolumn{2}{c}{87.35 $\pm$ 5.73} & \multicolumn{2}{c}{88.53 $\pm$ 2.12} & \multicolumn{2}{c}{95.58 $\pm$ 1.11} \\
Conv-branch  & 90.26 & 1.98 & \multicolumn{2}{c}{87.06 $\pm$ 6.41} & \multicolumn{2}{c}{88.35 $\pm$ 3.25} & \multicolumn{2}{c}{95.38 $\pm$ 1.32} \\
Dual-branch  & \textbf{92.23} & \textbf{1.22} & \multicolumn{2}{c}{\textbf{89.78 $\pm$ 9.97}} & \multicolumn{2}{c}{\textbf{90.02 $\pm$ 4.23}} & \multicolumn{2}{c}{\textbf{96.91 $\pm$ 1.91}} \\ \hline
\end{tabular}
}
\vspace{1ex}
\caption{Comparison of the two branches of SDAPC block on the ACDC dataset. SDMSA-branch strands for the model with only the Swin Deformable MSA branch in its basic block. Conv-branch stands for the model with only the convolution branch in its basic block. Dual branch stands for the proposed model that combines both the Swin Deformable MSA branch and the convolution branch in its basic block.}
\label{tab:ablation two branch}
\vspace{-2ex}
\end{table}

\vspace{0ex}
\section{Conclusions}

In this study, we have developed the Swin Deformable MSA with Parallel Convolution block, i.e., SDAPC block, and proposed the Swin Deformable Attention Hybrid UNet, i.e., SDAH-UNet, for explainable medical image segmentation. The experiments have proven that SDAH-UNet outperforms the state-of-the-art model on both anatomy and lesion segmentation . The ablation studies have illustrated that both the SDMSA module and the Parallel Convolution layer improved the segmentation performance of SDAH-UNet. Moreover, the deformation points can illustrate two-level of explanation: the model focalization and how the model forms the focalization. The explanations have also proven that the SDAPC block enables the model to focus more precisely on the segmentation targets, reduce MSA computation redundancy, and leads to performance enhancement.

\newpage
\bibliographystyle{IEEEtran}
\bibliography{citation}

\end{document}